# Pressure effects on the transport coefficients of $Ba(Fe_{1-x}Co_x)_2As_2$

S. Arsenijević,[1] R. Gaál,[1] A. S. Sefat,[2] M. A. McGuire,[2] B. C. Sales,[2] D. Mandrus,[2,3] and L. Forró[1]

[1]Institute of Condensed Matter Physics, Swiss Federal Institute of Technology, EPFL, CH-1015 Lausanne, Switzerland

[2]Materials Science and Technology Division, Oak Ridge National Laboratory, Oak Ridge, TN 37831, USA

[3]Department of Materials Science and Engineering, University of Tennessee, Knoxville, TN 37996, USA

## Abstract

We report the temperature dependence of the resistivity and thermoelectric power under hydrostatic pressure of the itinerant antiferromagnet $BaFe_2As_2$ and the electron-doped superconductor $Ba(Fe_{0.9}Co_{0.1})_2As_2$. We observe a hole-like contribution to the thermopower below the structural-magnetic transition in the parent compound that is suppressed in magnitude and temperature with pressure. Pressure increases the contribution of electrons to transport in both the doped and undoped compound. In the 10% Co-doped sample, we used a two-band model for thermopower to estimate the carrier concentrations and determine the effect of pressure on the band structure.

PACS numbers: 74.70.Xa, 74.25.F, 74.25.fg

## Introduction

The recently discovered Fe-based superconductors with the maximum $T_C$ of 55K, have evoked great interest, chiefly because of their many similarities to cuprate high $T_C$ perovskites: the ground state is antiferromagnetic, doping is needed to induce superconductivity in the mother compound, and in many of these materials a conducting layer is responsible for superconductivity. However, the differences are also numerous: undoped Fe-pnictides are semimetals with all five Fe-3d bands crossing the Fermi level. This gives rise to 3 hole and 2 electron sheets. Nesting between them leads to a spin density wave with partial gapping of the Fermi-surface. The fact that these materials are semimetallic with the top of the valence and the bottom of the conduction band in the proximity of the Fermi energy makes them very sensitive to small changes of external parameters like doping or pressure. The doping or pressure shift the balance between electrons and holes, leading to changes of the Fermi surface that modify the nesting condition and this is suggested to be the mechanism for the suppression of antiferromagnetism and the emergence of superconductivity.[1] Similarities between the effect of doping and pressure were reported for structural changes in Fe-pnictides by Kimber *et al*., and confirmed by DFT calculation.[2]

For the description of the electronic properties of certain Fe-pnictides, Sales *et al.*[3] employed a simplified two-band model, considering only one electron and one hole band. It describes successfully the nontrivial temperature dependence of the magnetic susceptibility, the Hall effect and the Seebeck coefficient as a function of doping. Our motivation is to check the applicability of this two-band model for the pressure dependence of the electrical transport in $Ba(Fe_{1-x}Co_x)_2As_2$. The study of electrical transport properties and their evolution with pressure give us a perfect tool to investigate indirectly the Fermi surface and the changes of the electron system and then to compare them to the changes induced by doping. Thermoelectric power is a particularly useful tool to study the change of the relative weight of holes and electrons since their contribution is of opposite sign to the total thermopower, thus any change in the balance shows up much more clearly than in the resistivity. We used the two-band model from Ref. [3] to model the pressure-induced changes of the band structure in Co-doped $BaFe_2As_2$ and compare the obtained results with the reported Co-doping induced changes of the Fermi surface seen by ARPES.[4]

## Experimental methods

$BaFe_2As_2$ and $Ba(Fe_{1-x}Co_x)_2As_2$ single crystals were grown using the FeAs self-flux method.[5] Pieces of typical dimensions of 2mm x 0.5mm x 0.1mm were cut and cleaved from single crystals for transport measurements. The sample was thermally anchored to a ceramic holder on which a variable thermal gradient was generated by a heater positioned at one end. This gradient was measured with a differential thermocouple. Sample contacts were made by gold wires with silver epoxy for four-lead resistivity and thermoelectric power measurements. The whole setup fits in the volume of a clamped piston-cylinder pressure cell. Hydrostatic pressure up to 2.5GPa was achieved by using Daphne oil 7373 as a pressure medium. The contribution of the gold leads to the thermopower is determined to be much smaller than that of the sample, thus no correction for it has been applied.

## Results

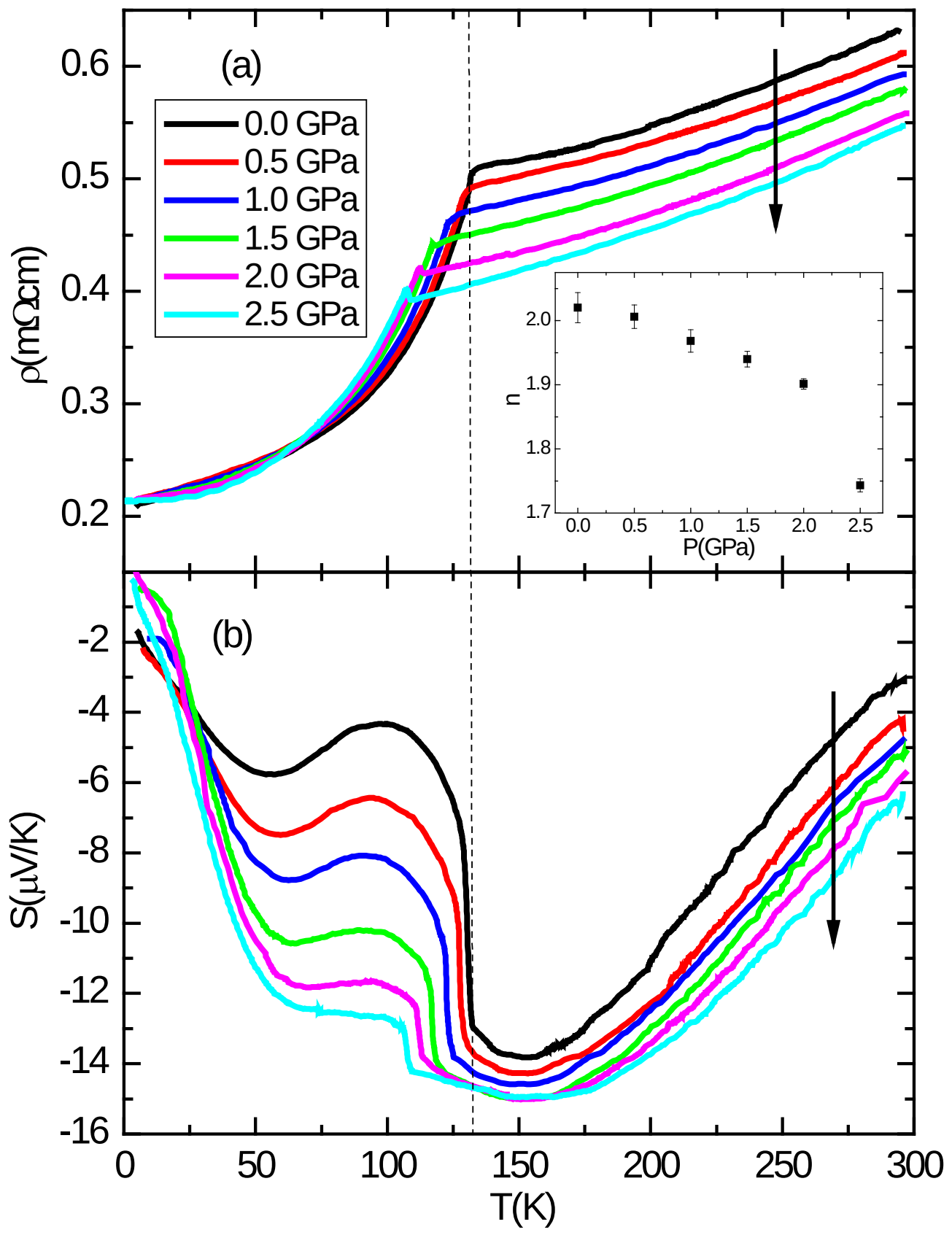

FIG. 1. (a) Resistivity under pressure for the parent compound $BaFe_2As_2$. The resistivity above 150K was fitted with a power law of the form $A+BT^n$, and the inset shows the pressure dependence of the exponent *n*. The monotonic decrease of *n* suggests an increase of the weight of the electron band according to the Ref. [6] (b) Thermopower of the $BaFe_2As_2$ under pressure. The arrow indicates the variation of TEP with pressure. The vertical line that is drawn at the transition temperature of the ambient pressure curve shows that the sudden increase of the hole contribution to the TEP coincides with the drop in the resistivity curve.

In Fig. 1(a) we present the temperature dependence of the resistivity of the parent compound at different pressures. The concomitant structural/magnetic (S-M) transition at ambient pressure is signalled by a sudden drop of the resistivity at $T$ = 132K. With pressure the S-M transition temperature is suppressed and the two are separated above 1.5 GPa, as indicated by the step-like increase of resistivity at the transition. This is in full analogy with the behaviour upon Co doping.[7] The high temperature part of the curves seem to be shifted with respect to each other, suggesting a strong variation of the residual resistivity with pressure, while at low temperature all curves converge to the same point. This shows that the residual resistivity in the two temperature ranges has different origins. The suppression of the magnetic transition temperature goes hand in hand with the decrease of the residual resistivity in the high temperature part, hinting to magnetic fluctuations as charge scatterers. We have fitted a power law in the paramagnetic temperature regime to study the temperature dependence of the exponent. The inset in the figure shows that the exponent *n* is decreasing from 2 to 1.7.

Fig. 1(b) shows the thermoelectric power (TEP) *vs*. temperature, at the same pressures as the resistivity. The TEP of $BaFe_2As_2$ is negative in the whole temperature range and it attains its maximum absolute value around 150K. The S-M transition manifests itself in a dramatic drop of the electron contribution to the TEP. Pressure makes the TEP more negative over the whole temperature range indicating the relative increase of the electron contribution. The magnitude of the effect of the S-M transition decreases with pressure, but it is still present even at the highest pressure of our study. At higher pressures, where the structural and magnetic transitions are separated (indicated by the small peaks in the resistivity in Fig. 1(a), it becomes clear that the steep increase of the hole contribution is related to the structural transition, and not to the magnetic one.

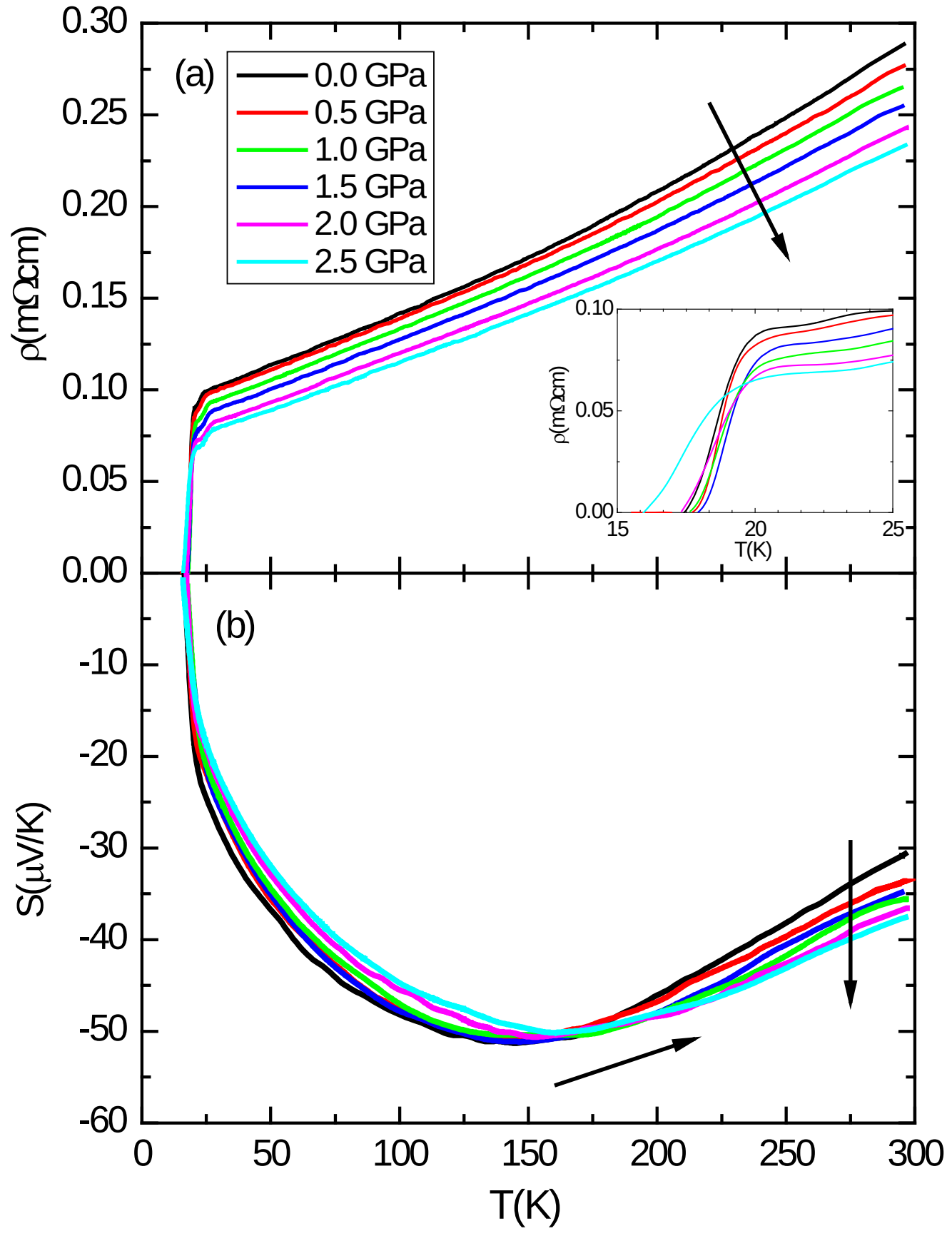

FIG. 2. (a) Resistivity vs. temperature of the superconducting $Ba(Fe_{0.9}Co_{0.1})As_2$ sample at different pressures. The inset shows a zoom on the superconducting transition. (b) Thermopower of the 10% Co doped sample under pressure. The arrows indicate the effect of pressure on the position of the minimum and the room temperature value.

In Fig. 2(a) we show the resistivity of the 10% doped compound. The room temperature resistivity decreases by a factor of two with respect to the undoped sample. There is no sign of the S-M transition, and superconductivity occurs below 19K. The critical temperature has a maximum at 1 GPa, but the overall pressure dependence of it is weak. The width of the transition however increases monotonically with pressure (inset in Fig. 2(a)). Resistivity below 100K is almost linear while the high-T dependence satisfies a power law with exponent $n \approx 1.4$.

The Seebeck coefficient (Fig. 2(b)) shows bigger change with doping: its magnitude in general increases by a factor of 5 while the general trend is similar to that of the undoped sample under highest pressure. With pressure the room temperature thermopower is increasing, and the maximum of $|S|$ is shifted from around 150K to higher temperatures while it slightly decreases in magnitude. Similar behavior of $S$ is observed with additional Co doping, from $x$=0.1 to $x$=0.12.[8] At the superconducting transition the thermopower drops to zero.

Similarly high thermoelectric power with a maximum in the 200-400K temperature range has been found in underdoped cuprates (see [9] and references cited therein), where the maximum in temperature has been attributed to a transition from coherent to incoherent excitations, a consequence of strong correlations.[10] In both families, the cuprates and Fe-based superconductors, the topology of the Fermi surface changes, [4, 11] Lifshitz-transitions take place,[12] to which thermopower is extremely sensitive.[13,14]

## Interpretation and modeling

Both experimental results and theoretical calculations show that in order to describe the transport properties of pnictides one has to employ at least a two-band model in order to account for competing electron and hole bands present at the Fermi level. Hall effect measurements have shown that if holes are the dominant charge carriers, the resistivity varies as $T^2$, whereas electron-dominated conduction leads to nearly linear $T$-dependence.[6] The decrease of the power law exponent we observe with pressure in the undoped compound or with doping indicates that the weight of the electron band is increased both by pressure and doping.

Looking at the thermopower and the resistivity curves at the same time allows the following qualitative statements. The SDW gap formation at the AFM transition does not involve all the bands since the resistivity is decreasing below the transition temperature. Upon separation of the structural and magnetic transitions above 2.0 GPa we can separate the effect of the two transitions. The higher temperature structural change corresponds to the increase of resistivity and to a high hole-like positive signature in thermopower. We can ascribe this to the reconfiguration of electronic structure with an appearance of a hole-like band. The lower temperature AFM/SDW transition is indicated by the decrease of resistivity and the vanishing positive effect in $S$ signalling a fall of the hole band below the $E_F$.

These changes in the electronic structure have been confirmed by recent ARPES measurements.[4]Those authors have observed the appearance of a petal-like hole pocket in the Brillouin zone corner $X$ below $T_{SDW}$. This pocket is the signature of the AFM transition and $x$=0.03 Co doping suppresses it below $E_F$, leading to the Lifshitz transition at the onset of superconductivity. Thermoelectric power measurements on samples with different Co-doping confirm that 3% doping is enough to completely eliminate the sudden positive contribution to thermopower.[8] The pressure-induced decrease of the hole contribution to the thermopower in the undoped sample is also a signature of a loss of weight of this hole pocket, thus indicating similar effect of pressure and electron doping. At our maximal pressure the trace of this pocket is still visible and the fact that we do not observe SC is coherent with previous results.

For deeper understanding of the variations of the band structure we used a two-band model to analyse our results.[3,15] In this model the total thermopower is the weighted sum of the contributions from the two bands, the weights being the respective conductivities. The strong electron-hole interaction, which leads to the SDW transition in the parent compound, justifies the choice that interband scattering is supposed to be the dominant relaxation mechanism, leading to the same relaxation time for both carriers. This gives the following expression for the thermopower:

$$S = \frac{S_e\sigma_e + S_h\sigma_h}{\sigma_e + \sigma_h} \approx \frac{S_e N / m_e^* + S_h P / m_h^*}{N / m_e^* + P / m_h^*}. \quad (1)$$

The electron and hole bands are taken to be parabolic, with fitting parameters being the effective masses of the charge carriers ($m_e$, $m_h$) and the band overlap ($E_h$). The rigid band scenario[16] justifies that the three band-related parameters should be kept constant, while the only parameter changing with doping is the excess electron concentration $N_0$, which increases linearly with Co content. From the expressions for the number of electrons ($N$) and holes ($P$) for semimetals and the charge balance constraint[3,15, 17], $N = P + N_0$ (where $N_0$ represents the excess electron concentration at $T$=0), one can determine the chemical potential $E_F$ at any given temperature. This allows us to calculate $S_e$ and $S_h$ assuming simple metal-like diffusion thermopower for the two bands, and adding them. With the help of this model, Sales *et al.*[3] successfully described the temperature dependence of the magnetic susceptibility, Hall effect and the thermopower at different doping levels.

From the 4 data sets (resistivity and TEP of the two compounds), in our detailed analysis we concentrate on the TEP of the doped sample. The reason is twofold: first, electrons and holes contribute to TEP with different signs, so change in their relative weight will have bigger effect than in resistivity. Secondly, in the case of the doped sample the S-M transition is absent; in other terms the electron structure can be assumed to be the same from room temperature down to the superconducting transition, giving us a wide temperature range to be analyzed.

Pressure can influence band structure most directly by increasing the bandwidth, which implies changes in the carrier effective masses and band overlaps. Thus, out of the four parameters of the model, first we tried to allow pressure dependence of $m_e$, $m_h$ and $E_h$. However, we found non-monotonic variation of the parameters with pressure and very wide confidence intervals. Since we believed that this is mostly due to strong correlation between the parameters, we decided to proceed to a systematic statistical analysis in order to select the minimum number of pressure dependent parameters to give a satisfactory description of the data. As a result of this approach we have found that fixed effective masses and pressure-dependent $E_h$ and $N_0$ yield satisfactory fits. The agreement between the modeled and measured thermopower is excellent (Fig. 3(a)), and the pressure dependence of the two parameters is essentially linear (inset in Fig. 3(a)). Our approach can also account for the effect of Co doping in this compound,[8] explaining the simultaneous decrease of $|S|_{max}$ above $x$=0.06 and the increase of $|S|$ at room temperature with $x$ better than the original model which allows variation only of $N_0$ with doping,[3] thus proving the similar effect of pressure and doping. The pressure independent effective masses used in the fit are $m_e$=22.75 and $m_h$=17.25. The high values of the effective masses which have already been discussed in Ref. [3] are a signature of the correlated character of the d-band, and are also being expected form the modestly high Sommerfeld coefficient determined from heat capacity measurements[18,19] and recent De Haas-van Alphen measurements on similar compounds.[20]

For a deeper insight, we have calculated the carrier densities as a function of temperature at different pressures and we have found that the hole concentration is independent of pressure while the electron concentration increases considerably (Fig. 3(b)). This is similar to the effect of Co doping seen in ARPES measurements,[4] *i.e.*, the observed increased intensity of the electron band. We believe that this is a key fact, and this has led us to choose the top of the hole band as the reference level when showing the variation of the band structure (Fig. 4). In this picture, the electron band is pushed deeper and deeper below the Fermi-level, where the $E_F$ does not change substantially its position with respect to the top of the hole band. The observed robustness of the hole band and $T_C$ with applied pressure again parallels the behavior with Co doping where it was shown that the existence of the hole-like Γ Fermi sheet is a necessary condition for SC,[4] since it is believed that the pairing coupling comes from the inter-band interaction.[21] Although our calculations indicate that the number of hole carriers $P$ diminishes close to $T_C$ (Fig. 3(b)), a small change of the values of parameters $m_e$ and $m_h$ would yield a finite value of $P$ at $T$=0K.

For the electronic transport our scenario has the following consequences. At very low temperature, the Fermi-level lies well within the electron band, electrons are the majority carriers, and the carrier concentration does not have a strong temperature dependence. In agreement with this, the thermopower is negative and linear in $T$. Furthermore, in accordance with the results of Ref [22], the resistivity is also linear. Upon increasing temperature, the Fermi-level moves downwards, as dictated by carrier conservation, and at some point it drops below the top of the hole band (Fig. 4(a)). The concentration of both carriers increases, and the appearance of holes leads to a positive contribution to the thermopower. However, their contribution to the transport in the doped compound is much smaller compared to that of electrons and it is only important in the temperature region above 100K.

The best description of the pressure dependence of the thermopower of the doped compound is obtained if some variation of the doping level with pressure is allowed, which is not straightforward to understand. Qualitative explanation might come from the results of Wadati, Elfimov and Sawatzky,[23] where the authors, using a DFT modelling, found strong localization of the additional electrons introduced by doping, and suggested an alternative scenario explaining how doping induces superconductivity. It is well possible that pressure helps to delocalise these extra electrons, hence the change of $N_0$.

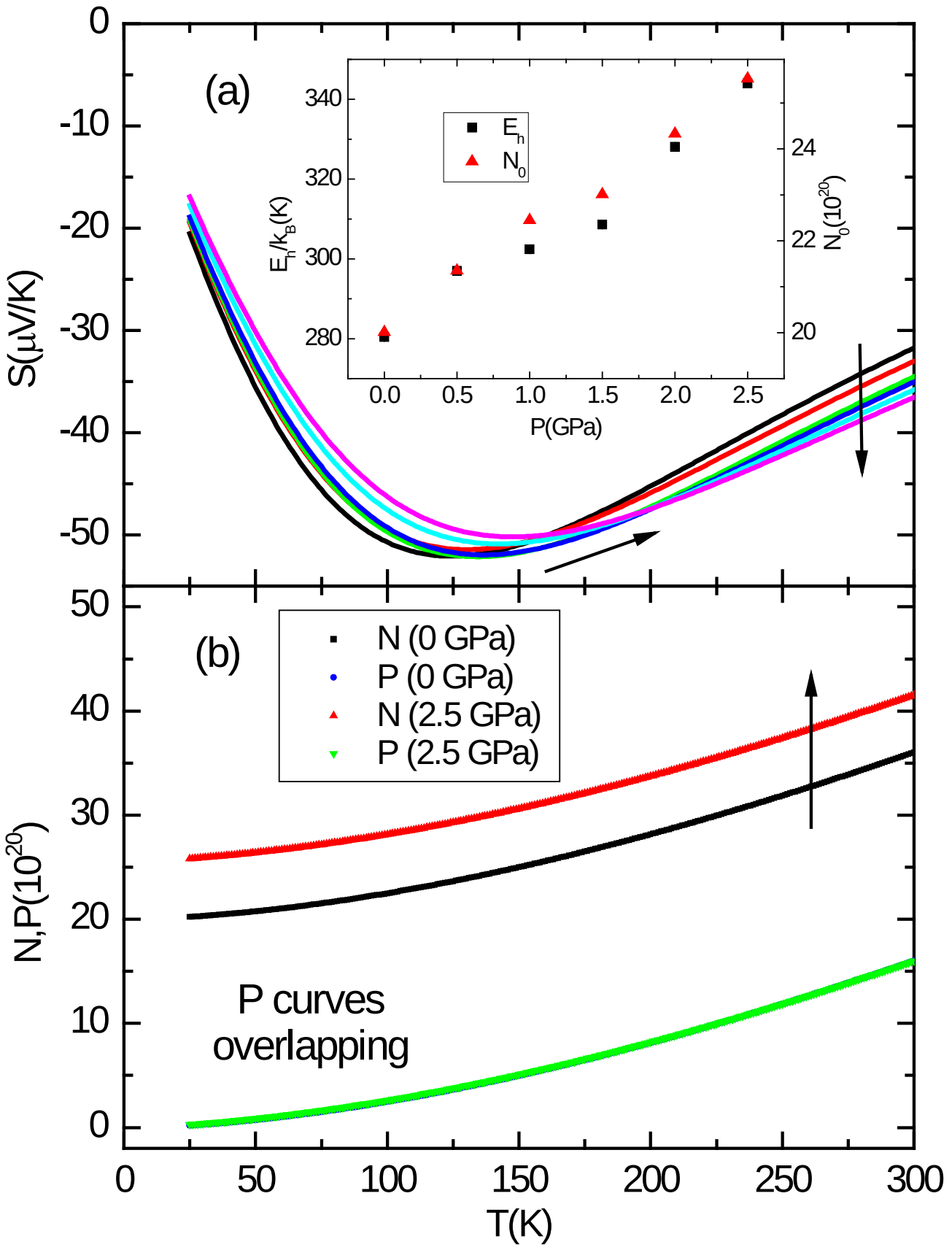

FIG. 3. (a) Modelled thermopower based on a fit to our thermopower data presented in Fig. 2(b). The arrows indicate the change with pressure of some characteristic features similar to the experimental results. The inset shows the pressure dependence of the model parameters: the number of carriers at zero temperature $N_0$ and the electron-hole band overlap $E_h$. (b) Temperature dependence of electron (N) and hole (P) concentration at ambient and maximal applied pressure. While the hole concentration is nearly pressure independent, the electron concentration increases with applied pressure.

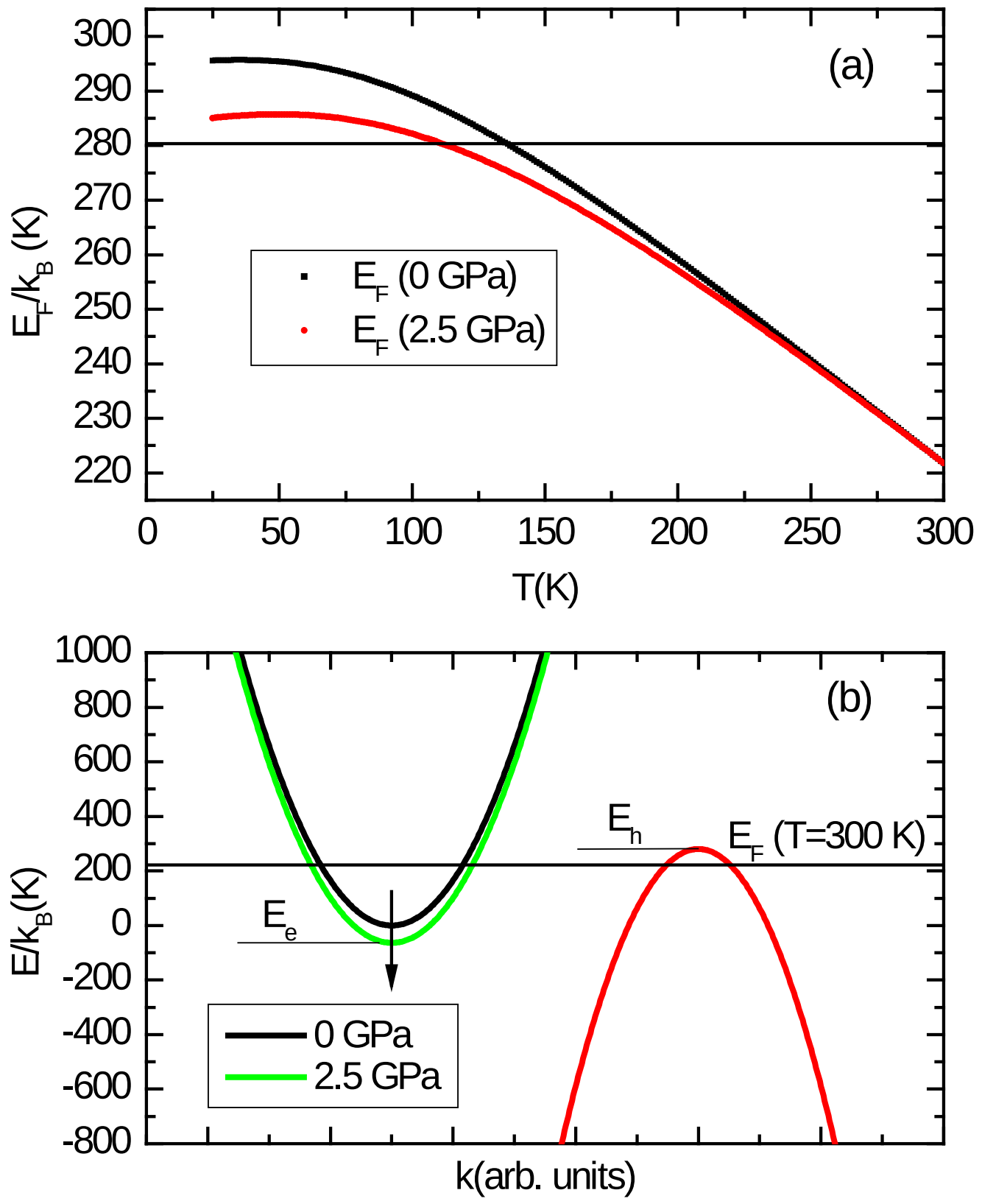

FIG. 4. (a) Position of the Fermi energy ($E_F$) with respect to the top of the hole band. (b) The evolution of the band structure with pressure according to model calculations.

As a result, the electron Fermi surface increases while the hole surface remains constant or becomes smaller. The ratio between the electron and hole surfaces is increased indicating that the electrons become the dominant carriers in the system. The effect of pressure is in an excellent agreement with the effect of doping observed by ARPES[4] and Hall data.[22]

## Conclusion

We have measured the resistivity and the thermoelectric power as a function of temperature at different pressures in $BaFe_2As_2$ and $Ba(Fe_{0.9}Co_{0.1})_2As_2$. Thermopower is a transport coefficient that is very sensitive to changes of the Fermi-surface. A simplified two band model gives a satisfactory explanation of the physics in the semimetal Fe-pnictides compounds. Within this model we observed the increasing influence of electrons in transport with increasing pressure. With pressure and/or doping metallicity gets more pronounced, especially at low temperatures. This is in agreement with the effect observed with doping indicating that pressure and doping induce similar changes in the Fermi-surface.

## Acknowledgment

This work has been supported by the Swiss NSF and by the MaNEP NCCR. Research at ORNL was supported by the U.S. Department of Energy, Office of Basic Energy Sciences, Materials Sciences and Engineering Division.

## References

1. I. I. Mazin, D. J. Singh, M. D. Johannes and M. H. Du, Physical Review Letters **101**, 057003 (2008).
2. S. A. J. Kimber, A. Kreyssig, Y.-Z. Zhang, H. O. Jeschke, R. Valentí, F. Yokaichiya, E. Colombier, J. Yan, T. C. Hansen, T. Chatterji, R. J. McQueeney, P. C. Canfield, A. I. Goldman and D. N. Argyriou, Nature Materials **8**, 471-475 (2009).
3. B. C. Sales, M. A. McGuire, A. S. Sefat and D. Mandrus, Physica C: Superconductivity **470**, 304-308 (2010).
4. C. Liu, T. Kondo, R. M. Fernandes, A. D. Palczewski, E. D. Mun, N. Ni, A. N. Thaler, A. Bostwick, E. Rotenberg, J. Schmalian, S. L. Bud'ko, P. C. Canfield and A. Kaminski, Nature Physics **6**, 419-423 (2010).
5. A. S. Sefat, R. Jin, M. A. McGuire, B. C. Sales, D. J. Singh and D. Mandrus, Physical Review Letters **101**, 117004 (2008).
6. F. Rullier-Albenque, D. Colson, A. Forget, P. Thuéry and S. Poissonnet, Physical Review B **81**, 224503 (2010).
7. J.-H. Chu, J. Analytis, C. Kucharczyk and I. Fisher, Physical Review B **79**, 014506 (2009).
8. E. Mun, S. Bud’ko, N. Ni, A. Thaler and P. Canfield, Physical Review B **80**, 054517 (2009).
9. Y. Horiuchi, W. Tamura, T. Fujii and I. Terasaki, Superconductor Science and Technology **23**, 065018 (2010).
10. J. Merino and R. H. McKenzie, Physical Review B **61**, 7996–8008 (2000).
11. D. LeBoeuf, N. Doiron-Leyraud, B. Vignolle, M. Sutherland, B. Ramshaw, J. Levallois, R. Daou, F. Laliberté, O. Cyr-Choinière, J. Chang, Y. Jo, L. Balicas, R. Liang, D. Bonn, W. Hardy, C. Proust and L. Taillefer, Physical Review B **83**, 054506 (2011).
12. I. M. Lifshitz, Zh. Eksp. Teor. Fiz. **38**, 1569 (1960).
13. J. Lin, Physical Review B **82**, 195110 (2010).
14. Y. M. Blanter, M. I. Kaganov, A. V. Pantsulaya and A. A. Varlamov, Physics Reports **245**, 159 (1994).
15. H. J. Goldsmid, *Electronic Refrigeration*. (Pion Limited, London, 1986).
16. M. Neupane, P. Richard, Y. M. Xu, K. Nakayama, T. Sato, T. Takahashi, A. V. Federov, G. Xu, X. Dai, Z. Fang, Z. Wang, G. F. Chen, N. L. Wang, H. H. Wen and H. Ding, Physical Review B **83**, 094522 (2011).
17. S. M. Sze, *Physics of Semiconductors Devices*. (John Wiley and Sons, New York, 1981).
18. K. Gofryk, A. S. Sefat, E. D. Bauer, M. A. McGuire, B. C. Sales, D. Mandrus, J. D. Thompson and F. Ronning, New Journal of Physics **12**, 023006 (2010).
19. F. Hardy, T. Wolf, R. A. Fisher, R. Eder, P. Schweiss, P. Adelmann, H. v. Löhneysen and C. Meingast, Physical Review B **81**, 060501(R) (2010).
20. T. Terashima, M. Kimata, N. Kurita, H. Satsukawa, A. Harada, K. Hazama, M. Imai, A. Sato, K. Kihou, C.-H. Lee, H. Kito, H. Eisaki, A. Iyo, T. Saito, H. Fukazawa, Y. Kohori, H. Harima and S. Uji, Journal of the Physical Society of Japan **79**, 053702 (2010).
21. R. M. Fernandes and J. Schmalian, Physical Review B **82**, 014521 (2010).

22. F. Rullier-Albenque, D. Colson, A. Forget and H. Alloul, Physical Review Letters **103**, 057001 (2009).
23. H. Wadati, I. Elfimov and G. Sawatzky, Physical Review Letters **105**, 157004 (2010).